\documentclass[aps,pra,letterpaper,twocolumn,showpacs,superscriptaddress,floatfix,longbibliography]{revtex4-1}
\usepackage[english]{babel}
\usepackage{amsmath}
\usepackage{amsfonts}
\usepackage{amssymb}
\usepackage{color}
\usepackage{epsfig}
\usepackage{graphicx}
\usepackage{bm}
\usepackage{mathtools}
\usepackage{bbold}
\usepackage{times}
\usepackage{cancel}
\usepackage[colorlinks,urlcolor=blue,bookmarks=false,hypertexnames=true]{hyperref} 
\usepackage{flushend}

\allowdisplaybreaks

\newcommand{\bra}[1]{\langle #1 |} \newcommand{\ket}[1]{| #1 \rangle}
\newcommand{\scp}[2]{\langle #1 | #2 \rangle}
\newcommand{\braket}[3]{\langle #1 | #2 | #3 \rangle}

\newcommand{\Span}{\mathrm{span}} %
 
\newcommand{\tr}{\mathop{\mathrm{tr}}\limits}
\newcommand{\trcond}{\mathop{\mathrm{tr}_\mathrm{cond}}\limits}
\newcommand{\trnorm}{\mathop{\mathrm{tr}_\mathrm{norm}}\limits}
\newcommand{\Np}{N_{p}}
\newcommand{\Nimp}{N_{i}}
\newcommand{\gc}{g_{c}}
\newcommand{\Tc}{T_{c}}
\newcommand{\Gammac}{\Gamma_{c}}

\definecolor{myred}{RGB}{168,5,14}
\definecolor{myblue}{RGB}{13,13,255}
\definecolor{mylightblue}{RGB}{120,120,255}
\definecolor{mygreen}{RGB}{20,150,20}
\definecolor{mygray}{gray}{0.5}
\definecolor{myblack}{RGB}{0,0,0}
\definecolor{violet}{RGB}{200,0,200}

\definecolor{cpcolor}{RGB}{12,30,255}

\begin{document}

\title{Finite-temperature quantum condensations in the space of
  states: A new perspective on quantum annealing}

\author{Massimo Ostilli} \affiliation{Instituto de F\'isica,
  Universidade Federal da Bahia, Salvador, 40210-340, BA, Brazil}
\author{Carlo Presilla} \affiliation{Dipartimento di Matematica,
  Sapienza Universit\`a di Roma, Piazzale A. Moro 2, Roma 00185,
  Italy} \affiliation{Istituto Nazionale di Fisica Nucleare, Sezione
  di Roma 1, Roma 00185, Italy}

\date{\today}

\begin{abstract}
  In nature, everything occurs at finite temperature and quantum phase
  transitions (QPTs) cannot be an exception.  Nevertheless, they are
  still mainly discussed and formulated at zero temperature.  We show
  that the condensation QPTs recently introduced at zero temperature
  can naturally be extended to finite temperature just by replacing
  ground state energies with corresponding free energies.  We
  illustrate this criterion in the paradigmatic Grover model and in a
  system of free fermions in a one-dimensional inhomogeneous lattice.
  In agreement with expected universal features, the two systems show
  structurally similar phase diagrams. Last, we explain how finite
  temperature condensation QPTs can be used to construct quantum
  annealers having, at finite temperature, output-probability
  exponentially close to 1 in the system size. As examples we consider
  again the Grover model and the fermionic system, the latter being
  well within the reach of present heterostructure technology.
\end{abstract}


\maketitle

\section{Introduction}%
Quantum phase transitions (QPTs), i.e, the thermodynamic singularities
emerging at zero temperature ($T=0$) driven by some Hamiltonian
parameter of the system, originate from quantum fluctuations, a
consequence of Heisenberg's uncertainty principle.  However, an
isolated system at $T=0$ represents an abstract limit and
understanding the finite temperature counterpart of a QPT (if any) is
of paramount importance.  Such an aim represents a quite challenging
issue, from both the theoretical and experimental viewpoints since,
above zero temperature, quantum and thermal fluctuations may compete
in an intricate manner.

Put in simple terms, in general, two kinds of scenarios are
expected~\cite{SGCS,KB,Vojta,Sachdev,Zanardi2007,Fidelity,Carr,Plastino}:
the ordered phase exists only at zero temperature; the ordered phase
exists also at finite temperature, below some critical value $\Tc$
which, in turn, might signal a purely classical phase transition when
the Hamiltonian parameters are set to a value rendering the system
classical (e.g., the Ising model in a transverse field becomes the
classical Ising model when the transverse field is set to zero).

This second scenario is particularly appealing for potential
applications to quantum annealing protocols aimed at finding the
ground state (GS) by working at finite temperature, the GS of the
ordered phase being the solution of some (classical) combinatorial
problem of interest, possibly hard~\cite{Sebenik,Nishimori,Santoro}.
First-order QPTs, a much less explored field when compared to
second-order QPTs, might favor this scenario since, at the transition,
the order parameter jumps between two very different quantum states,
possibly orthogonal.

For concreteness, let us consider systems described by a Hamiltonian
of the form
\begin{eqnarray}
  \label{H}
  {H}=\Gamma {K}+J{V},
\end{eqnarray}
where $K$ and $V$ are two noncommuting dimensionless Hermitian
operators, and $\Gamma$ and $J$ are parameters with energy dimensions.
Representing $H$ in the eigenbasis of $V$, it is natural to call $V$
the potential operator, $K$ the hopping operator, $\Gamma$ the hopping
parameter and $J$ the potential strength.  We will use, equivalently,
$\Gamma$ or $J$ as the control parameter of the supposed QPT.  Since
phase transitions occur in the thermodynamic limit (TDL), we need a
fair competition between $K$ and $V$ in this limit.  By this we mean
that, supposing that $H$ describes a system of $N$ particles or spins,
the eigenvalues of $K$ and $V$ both scale linearly with $N$.

Recently, we have introduced a class of first-order QPTs taking place,
at $T=0$, via a condensation in the space of states~\cite{QPTA}.
Consider a partition of the space of states $\mathcal{H}$ of the
system into two subspaces,
$\mathcal{H} = \mathcal{H}_\mathrm{cond} \oplus
\mathcal{H}_\mathrm{norm}$, such that, in the TDL,
$\dim \mathcal{H}_\mathrm{cond}/ \dim \mathcal{H} \to 0$ and suppose
to follow, for a very large time $t$, the evolution of an initial
quantum state belonging to $\mathcal{H}_\mathrm{cond}$.  A heuristic,
ergodicity-breaking argument, applies, which is inspired by the exact
probabilistic representation of the quantum evolution introduced
in~\cite{BPDAJL}:
due to the fact that $\mathcal{H}_\mathrm{cond}$ has vanishing
relative dimension, in the TDL, the system will spend a vanishing
fraction of $t$ inside $\mathcal{H}_\mathrm{cond}$ unless it finds
that it is energetically more convenient to remain in
$\mathcal{H}_\mathrm{cond}$.  In other words, at $T=0$, in the TDL the
energy of the system becomes the minimum between the GS energies of
$H$ restricted to $\mathcal{H}_\mathrm{cond}$ and
$\mathcal{H}_\mathrm{norm}$.  Now, if for some finite value of the
Hamiltonian parameter $\Gamma$ (or $J$), the above two GS energies
cross each other, the system undergoes a first-order QPT driven by
this parameter.  More precisely, crossing this point the GS of the
system steeply changes from being a superposition in the
\textit{normal} subspace $\mathcal{H}_\mathrm{norm}$, asymptotically
as large as $\mathcal{H}$, to being a superposition in the
\textit{condensed} subspace $\mathcal{H}_\mathrm{cond}$, with
vanishing relative dimension. This is what we call a condensation
QPT. Notice that, by construction, the \textit{normal} and
\textit{condensed} GS are orthogonal.
 
In the present paper, we propose a simple criterion to extend the
$T=0$ condensation QPTs to finite temperature.  For systems at
canonical equilibrium at temperature $T$, the same phase transition
mechanism described above can take place just by replacing the GS
energies with the free energies. While a mathematical proof of this
criterion can be provided~\cite{TH_QPT_proof}, here we illustrate it
with two examples.  First, we derive analytically the phase diagram of
the paradigmatic Grover model.  Then, we consider a physical system,
experimentally implementable, consisting of free fermions in a
one-dimensional (1D) inhomogeneous lattice.  In the latter case we
obtain, numerically, a phase diagram, that is structurally similar to
that of the Grover model, in agreement with the universal features of
the condensation QPTs~\cite{TH_QPT_proof}.  Guided by the above two
case studies, we finally show how finite-temperature condensation QPTs
can be used to build quantum annealers having, at finite temperature,
output-probability exponentially close to 1 in the system size.

Two comments are in order about the nature of the condensation QPTs:
(i) they are intrinsically first-order, for they can be driven by
using even one single Hamiltonian parameter.  In contrast, as for the
classical case, jumps of the order parameter can result when crossing
the coexistence line of two different phases that originate from the
critical point of a second-order QPT.  Notice that, for such a
scenario to take place at zero temperature, the Hamiltonian needs to
depend on at least two independent parameters (think of the 1D Ising
model in the presence of both transverse and longitudinal magnetic
fields~\cite{Continentino,Pelissetto}). (ii) Condensation QPTs are far
from being exotic.  As we recently showed at $T=0$, the renowned
Wigner crystallization belongs to this class of QPTs~\cite{WC_QPT}.

\section{Normal and condensed subspaces}%
We start by defining a proper partition of the space of states.
Consider a system with Hamiltonian~(\ref{H}), and let
$\{ \ket{\bm{n}_k} \}_{k=1}^{M}$ be a complete orthonormal set of
eigenstates of $V$, called \textit{the configurations}:
$V \ket{\bm{n}_k} =V_k \ket{\bm{n}_k}$, $k=1,\dots,M$.  We assume
ordered potential values $V_1 \leq \dots \leq V_M$.  Given an integer
$M_\mathrm{cond}$ with $1\leq M_\mathrm{cond} <M$, we make a partition
of the set of the configurations as
$\{ \ket{\bm{n}_k} \}_{k=1}^{M}=\{
\ket{\bm{n}_k}\}_{k=1}^{M_\mathrm{cond}}\cup \{
\ket{\bm{n}_k}\}_{k=M_\mathrm{cond}+1}^{M}$.  Correspondingly, the
Hilbert space of the system,
$\mathcal{H} = \Span \{ \ket{\bm{n}_k} \}_{k=1}^{M}$, equipped with
the standard complex scalar product $\scp{u}{v}$, is decomposed as the
direct sum of two mutually orthogonal subspaces, denoted condensed and
normal,
$\mathcal{H}=\mathcal{H}_\mathrm{cond} \oplus
\mathcal{H}_\mathrm{norm}$, where
$\mathcal{H}_\mathrm{cond} = \Span
\{\ket{\bm{n}_k}\}_{k=1}^{M_\mathrm{cond}} $, and
$\mathcal{H}_\mathrm{norm} = \Span \{
\ket{\bm{n}_k}\}_{k=M_\mathrm{cond}+1}^{M}
=\mathcal{H}_\mathrm{cond}^\perp$.

\section{Finite temperature quantum condensations}%
We suppose that the system, in contact with a heat bath, is at
canonical equilibrium at temperature $T=1/(k_B \beta)$, that is, it is
in the state described by the Gibbs density matrix operator
$\rho=e^{-\beta H}/\tr e^{-\beta H}$.  We define the Gibbs free
energies associated with the spaces $\mathcal{H}$,
$\mathcal{H}_\mathrm{cond}$, and $\mathcal{H}_\mathrm{norm}$ as
\begin{align*}
  &e^{-\beta F}
    =
    \tr e^{-\beta H}
    =
    \sum_{\ket{\bm{n}}\in\mathcal{H}}
    \braket{\bm{n}}{e^{-\beta H}}{\bm{n}},
  \\
  & e^{-\beta F_\mathrm{cond}} 
    =
    \trcond e^{-\beta H_\mathrm{cond}} 
    =
    \sum_{\ket{\bm{n}}\in\mathcal{H}_\mathrm{cond}}
    \braket{\bm{n}}{e^{-\beta H_\mathrm{cond}}}{\bm{n}},
  \\
  & e^{-\beta F_\mathrm{norm}}
    =
    \trnorm e^{-\beta H_\mathrm{norm}} 
    =
    \sum_{\ket{\bm{n}}\in\mathcal{H}_\mathrm{norm}}
    \braket{\bm{n}}{e^{-\beta H_\mathrm{norm}}}{\bm{n}},
\end{align*}
where $H_\mathrm{cond}$ and $H_\mathrm{norm}$ are the restrictions of
$H$ to the condensed and normal subspaces.  In the representation of
the eigenstates of $V$, $H_\mathrm{cond}$ corresponds to a null matrix
except for the block
$\braket{\bm{n}_k}{H_\mathrm{cond}}{\bm{n}_{k'}} =
\braket{\bm{n}_k}{H}{\bm{n}_{k'}}$,
$k,k'=1,\dots,M_\mathrm{cond}$. Analogously, $H_\mathrm{norm}$
corresponds to a null matrix except for the block
$\braket{\bm{n}_{k}}{H_\mathrm{norm}}{\bm{n}_{k'}} =
\braket{\bm{n}_{k}}{H}{\bm{n}_{k'}}$,
$k,k'=M_\mathrm{cond}+1,\dots,M$. Note that
$H_\mathrm{cond}+ H_\mathrm{norm} \neq H$.  According to the scaling
properties assumed for $K$ and $V$, we have that the free energies
$F$, $F_\mathrm{cond}$ and $F_\mathrm{norm}$ increase linearly with
$N$ (at least in the TDL).

By replacing the energies $E$, $E_\mathrm{cond}$, and
$E_\mathrm{norm}$ with the corresponding free energies $F$,
$F_\mathrm{cond}$, and $F_\mathrm{norm}$, in analogy with the $T=0$
case~\cite{QPTA}, we find that, if $M_{\mathrm{cond}}/M\to 0$, then up
to $o(N)$ terms,
\begin{align}
  \label{TH_QPTR}
  F \simeq \left\{
  \begin{array}{ll}
    F_\mathrm{cond}, \qquad &\mbox{if } F_\mathrm{cond}<F_\mathrm{norm}, 
    \\ 
    F_\mathrm{norm}, \qquad &\mbox{if } F_\mathrm{norm}<F_\mathrm{cond}.
  \end{array} \right.
\end{align}
See appendix~\ref{proofs.appendix} for the proof.  The above criterion
provides an extension of the $T=0$ condensation QPTs to finite
temperature.  In fact, by varying some parameter of the Hamiltonian,
e.g., $\Gamma$ or $J$, and/or the temperature $T$, we obtain a QPT,
necessarily of first order, whenever a crossing takes place between
$F_\mathrm{cond}$ and $F_\mathrm{norm}$.

Provided that the above criterion holds true, in the TDL the space of
states splits at the quantum critical line defined by
\begin{align}
  \label{QPT0}
  \lim_{N \to\infty }\frac{F_\mathrm{cond}}{N}=
  \lim_{N \to\infty }\frac{F_\mathrm{norm}}{N},
\end{align}
In other words, also for finite $T$, in the TDL we have effectively
$e^{-\beta H}\to e^{-\beta (H_\mathrm{cond}+H_\mathrm{norm})}$.  This
allows us to define the following order parameter which definitely
classifies the condensation in the space of states as a first-order
phase transition
\begin{align}
  \label{pcond3}
  p_\mathrm{cond} = 
  \sum_{\ket{\bm{n}}\in\mathcal{H}_\mathrm{cond}}
  \braket{\bm{n}}{\rho}{\bm{n}}
  \simeq 
  \frac{1}{1+e^{-\beta(F_\mathrm{norm}-F_\mathrm{cond})}},
\end{align}
where the last expression is valid up to terms exponentially small in
$N$; see appendix \ref{proofs.appendix} for the proof.  In the TDL the
free energies diverge as $N$, and we have $p_\mathrm{cond}=1$ in the
condensed phase, where $F_\mathrm{cond}<F_\mathrm{norm}$, and
$p_\mathrm{cond}=0$ in the normal phase, where
$F_\mathrm{norm}<F_\mathrm{cond}$.  At the critical line separating
the two phases we have $p_\mathrm{cond}=1/2$.

Apart from the necessary condition $M_{\mathrm{cond}}/M\to 0$ in the
TDL, the size $M_\mathrm{cond}$ should be properly chosen so that
Eq.~(\ref{QPT0}) admits a solution; see \cite{WC_QPT} for a detailed
discussion at $T=0$.
In particular, $M_\mathrm{cond}$ cannot be smaller than the degeneracy
of the GS of $V$ but can be bigger, as in the second example we
consider.

\section{Grover model}%
We illustrate the mechanism of the finite temperature condensation in
the exactly solvable Grover model, a simple yet non trivial example
emulating a benchmark model for quantum
search~\cite{Grover,Farhi.Goldstone,Roland.Cerf,Jorg:2008,Jorg:2010}.
For this model, the space of states $\mathcal{H}$ can be identified
with the space spanned by the $M=2^N$ spin states indicated by
$\ket{\bm{n}} = \ket{n_1} \ket{n_2} \dots \ket{n_N}$, where
$\ket{n_i}=\ket{\pm}$ is an eigenstate of the Pauli matrix
$\sigma^z_i$ relative to the qubit $i=1,\dots,N$.  The potential is
$V=\sum_{\bm{n}} V_{\bm{n}} \ket{\bm{n}}\bra{\bm{n}}$, where
$V_{\bm{n}}=-J N \delta_{\bm{n},\bm{n}_1}$, $J>0$, and $\bm{n}_1$
represents the target of a totally unstructured (worst case scenario)
search.  In contrast, structured searches correspond to potentials
with a smooth minimum around the target and, therefore, benefit from
the application of gradient-descent-based methods like the Ising
model, in which, however, the corresponding QPTs are second-order.
Finally, the hopping operator ${K}$ of the Grover model is chosen to
be the sum of single-flip operators $ {K}= - \sum_{i=1}^N \sigma^x_i$.

The GS of $K$ is nondegenerate and we choose $M_\mathrm{cond}=1$.  It
follows that $-\beta F_\mathrm{cond}= -\beta V_1 = \beta JN$.  Up to
corrections exponentially small in $N$, the free energy of the normal
subspace coincides with that of the hopping operator $K$ whose levels
are $-\Gamma(N-2j)$, $j=0,\dots,N$, and have degeneracy
$N!/[j!(N-j)!]$,
\begin{align}
  e^{-\beta F_\mathrm{norm}} = \tr e^{-\beta K}
  = \sum_{j=0}^{N} \binom{N}{j} e^{-\beta[-\Gamma(N-2j)]},
\end{align}
which yields
$ -\beta F_\mathrm{norm} =N\ln \left[ 2\cosh(\beta\Gamma) \right]$.
The critical line defined by Eq.~(\ref{QPT0}) is thus
\begin{align}
  J = k_B T \ln\left[ 2\cosh(\Gamma/k_B T) \right], 
  \label{JcT_Grover}
\end{align}
which was also found in Ref.~\cite{Jorg:2010} via perturbation theory.
Note that Eq.~(\ref{JcT_Grover}) has a solution only for
$J \geq \Gamma$.  A parametric plot of the critical line is shown in
the inset of Fig.~\ref{phase_diagram_fermions_impurities}.  For any
fixed $\Gamma$, at high temperature Eq.~(\ref{JcT_Grover}) provides
the asymptotic slope $T = J/(k_B\ln 2)$, while at low temperature the
slope becomes infinite, in agreement with the quantum critical point
$J=\Gamma$ of the $T=0$ transition.

\section{Free fermions in a 1D inhomogeneous lattice}%
Finite temperature condensation QPTs can be observed in a variety of
physically relevant systems. Here, we study the case of $\Np$ spinless
fermions in a 1D lattice with $N\geq \Np$ sites.  Some sites, the
first consecutive $\Nimp < \Np$, for simplicity, differ from the
others by the presence of an attractive potential so that the
Hamiltonian of the system is
\begin{align}
  \label{FermionH}
  H=-\eta\sum_{l=1}^{N-1} ( c^\dag_l c_{l+1} + c^\dag_{l+1} c_l )
  -g\sum_{l=1}^{\Nimp}c^\dag_l c_l,
\end{align}
where $c_l$ is the fermionic annihilation operator on site $l$ and we
choose open boundary conditions. The hopping parameter $\eta$ and the
attractive strength $g$ are positive constants. Note that we are
considering a system of noninteracting particles, nevertheless $H$ is
the sum of two noncommuting operators as in Eq.~(\ref{H}) and we can
look for a condensation QPT by varying the parameter $g$.

First, we define the subspace $\mathcal{H}_\mathrm{cond}$.  Besides
satisfying the necessary condition $M_{\mathrm{cond}}/M\to 0$ in the
TDL, this subspace should be large enough for the free energies
restricted to $\mathcal{H}_\mathrm{cond}$ and
$\mathcal{H}_\mathrm{norm}$ to cross each other at some finite value
$\gc$ of the parameter $g$.  The latter condition is equivalent to
having $F_\mathrm{norm}(g=0)<F_\mathrm{cond}(g=0)$ and
$\lim_{g\to\infty} F_\mathrm{cond}(g)/g <
\lim_{g\to\infty}F_\mathrm{norm}(g)/g$, see
Appendix~\ref{fermions.appendix.1} for an analysis of this system in
the $T=0$ limit.  In the present system and for $\Nimp=\Np$, the two
inequalities above are satisfied if $M_\mathrm{cond} = 1 + \Np^2$,
i.e., if the subspace $\mathcal{H}_\mathrm{cond}$ consists of the GS
of $V$, in which all the $\Np$ fermions are in the $\Nimp$ attractive
sites, and, in addition, of the $\Np^2$ first excited states of $V$,
in which $\Np-1$ fermions are in the $\Nimp$ attractive sites and one
is in the remaining $N-\Nimp$ sites.  Other choices of
$M_\mathrm{cond}$ are possible, but they all lead to the same TDL.
Remarkably, in this system, the nature of $\mathcal{H}_\mathrm{cond}$
shows that a condensation in the space of states corresponds to an
actual space localization in the attractive sites.
\begin{figure}
  \begin{center}
    {\includegraphics[width=\columnwidth,clip]
      {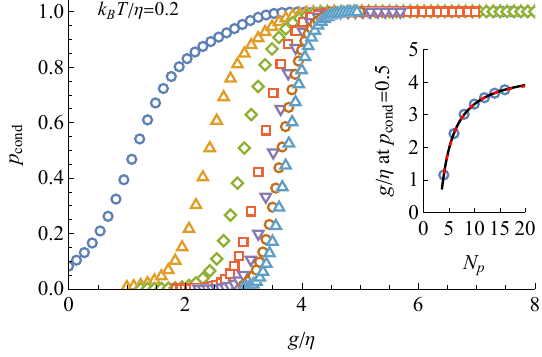}}
    \caption{Order parameter $p_\mathrm{cond}$ versus $g/\eta$ for the
      fermionic system~(\ref{FermionH}) with
      $\Np=\Nimp=N/2=4,6,8,10,12,14,16$, shown by symbols from left to
      right.  Here, $M_\mathrm{cond} = 1+\Np^2$, and the system is at
      canonical equilibrium at temperature $k_BT/\eta=0.2$. Inset:
      value of $g/\eta$ at $p_\mathrm{cond}=1/2$ for different $\Np$.
      Data (circles) compare quite well ($\chi^2 \simeq 10^{-4}$) with
      two fitting models $a+b/\Np+c/\Np^2$ (black solid line,
      $a=4.424$, $b=9.685$, and $c=13.760$) and
      $a+b \ln \Np +c \ln(\ln \Np )$ (red dashed line, $a=2.261$,
      $b=-2.268$, and $c=7.627$). The second model, however, must be
      rejected because in the TDL it gives $\gc\to -\infty$.  }
    \label{pcond_fermions_impurities}
  \end{center}
\end{figure}

Second, we evaluate the order parameter (\ref{pcond3}) as
\begin{align}
  p_\mathrm{cond} = \frac{1}{\sum_{j=1}^{M} e^{-\beta E_j}}
  \sum_{j=1}^{M} e^{-\beta E_j} 
  \sum_{k=1}^{M_\mathrm{cond}}
  |\scp{\bm{n}_k}{E_j}|^2,
  \label{pcond2}
\end{align}
where $E_j$ and $\ket{E_j}$ are the $\Np$-particle eigenvalues and
eigenvectors of $H$.  These are easily obtained with Pauli's principle
combining the single-particle eigenvalues and eigenvectors of $H$
calculated by numerically diagonalizing the $N\times N$ tridiagonal
matrix whose non zero elements are $A_{l+1,l}=A_{l,l+1}=-\eta$, for
$l=1,\dots,N-1$, and $A_{l,l}=-g$, for $l=1,\dots,\Nimp$.  The
computation of $E_j$ and $\ket{E_j}$ is a simple task which requires a
time $O(N^2)$. Also the computation of the sum of the squared scalar
products in Eq.~(\ref{pcond2}) requires an affordable time
$O(N^2\Np^2)$.  In fact, the evaluation of the scalar product between
the antisymmetrized states $\ket{\bm{n}_k}$ and $\ket{E_j}$ can be
reduced to the evaluation of the determinant of the $N\times N$ matrix
whose elements are the projections of the single-particle eigenvectors
of $H$ in the basis of the eigenvectors of $V$~\cite{Blaizot_Ripka}.
The calculation of this determinant requires a time $O(N^2)$.  It is
the sum over $M$ which limits the computation of $p_\mathrm{cond}$ to
a relatively small number of particles; in fact, $M$ grows as
$N!/[\Np! (N-\Np)!]$.

In Fig.~\ref{pcond_fermions_impurities} we show the behavior of
$p_\mathrm{cond}$ obtained as a function of the attractive strength
$g$ for a number of particles $\Np=4,6,8,\dots,16$.  As indicated
above, we chose $\Np=\Nimp=N/2$ and $M_\mathrm{cond} = 1+\Np^2$.  The
equilibrium temperature is $k_BT/\eta=0.2$, similar plots at different
temperatures are provided in Appendix~\ref{fermions.appendix.2}.
Whereas the order parameter exhibits a clear tendency toward the
step-like behavior expected for a first-order QPT, we are still far
from the TDL and no classical supercomputer would allow us to reach
much larger values of $\Np$.  However, we can estimate the critical
value $\gc$ of the QPT, namely, the value of $g$ at which
$p_\mathrm{cond}=1/2$ in the TDL, using a fit-and-extrapolate
procedure.  The validity of this procedure in the case of the Grover
model is illustrated in Appendix~\ref{Grover.appendix}.  We fit the
curve $a+b/\Np+c/\Np^2$ to the values of $g$ at which
$p_\mathrm{cond}=1/2$ for the available $\Np$ and extrapolate
$\gc=a$. A different fit model which provides a slowly diverging
$\gc$, namely, $a+b \ln \Np +c \ln(\ln \Np )$, must be rejected as we
obtain a pointless negative value of $b$, see the caption of
Fig.~\ref{pcond_fermions_impurities}.

\begin{figure}[t]
  \begin{center}
    {\includegraphics[width=\columnwidth,clip]
      {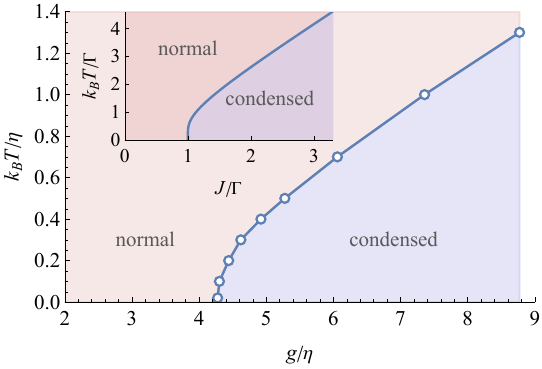}}
    \caption{ Phase diagram $g$-$T$ for the fermionic
      system~(\ref{FermionH}).  The solid line separating the two
      phases is obtained by connecting the dots evaluated by the
      fit-and-extrapolate procedure described in
      Fig.~\ref{pcond_fermions_impurities} for several values of
      $k_BT/\eta$. Inset: phase diagram $J$-$T$ for the Grover model,
      the solid line separating the two phases is drawn according to
      Eq.~(\ref{JcT_Grover}).}
    \label{phase_diagram_fermions_impurities}
  \end{center}
\end{figure}

In Fig.~\ref{phase_diagram_fermions_impurities} we plot the values of
$\gc$ derived as explained above for different temperatures.  The
result is a phase diagram in the $g$-$T$ plane having the same
universal features~\cite{TH_QPT_proof} as the phase diagram of the
Grover model. This corroborates the existence of a condensation QPT
for the present fermionic system.

The 1D fermionic system considered here can be investigated
experimentally by using superlattices, grown with reliable
technologies in the well established two-dimensional semiconductor
heterojunctions~\cite{2Dheterostructures} or in 1D
nanowires~\cite{nanowires}. In both cases, the alternation of
nano-layers of different materials forms effective 1D lattices made of
wells and barriers obtained by the corresponding band-gap energies.
By varying the barrier widths and heights one can tune the hopping
coefficients between neighboring wells, and by using different/doped
materials one can adjust the bottom level of the wells, i.e., create
local attractive potentials.  The number of electrons in the nanocells
of the superlattice can be fixed by photoexcitation, and their
dislocation at thermal equilibrium can be detected by
photoluminescence spectroscopy~\cite{Polimeni}. In this way, one can
have direct access to the size dependent order parameter and, for
large superlattices in which the TDL has been effectively reached, to
the critical line of the phase diagram.

\section{Condensation QPTs as efficient quantum annealers}%
Quantum annealers~\cite{Dwave} are physical devices generally aimed at
exploiting the quantum adiabatic theorem~\cite{Kato} to solve several
classes of optimization problems~\cite{Farhi.Gutmann}: by gradually
reducing $\Gamma$ an initial disordered quantum state of the system,
e.g., the GS of $K$ in Eq.~(\ref{H}), is made to evolve toward the
desired GS of $V$ in a time $\tau$ roughly given by the inverse of the
minimal gap of $H$.  In fact, the adiabatic condition is usually
formulated as $\tau=O(\Delta^{-2})$, with $\Delta$ being the minimal
gap of $H$. However, for models in which the potential $V$ is
non-degenerate, the weaker condition $\tau=O(\Delta^{-1})$ turns out
to be sufficient~\cite{Farhi.Goldstone,Warzel2015}.  Therefore, the
presence of a first-order QPT at $\Gamma_\mathrm{c}$ implies that
$\tau$ may grow exponentially with
$N$~\cite{Farhi.Goldstone,Jorg:2008,Jorg:2010,Altshuler,Warzel2015,ANN}.
This is the case of the Grover model, in which the minimal gap is
$2J N 2^{-N/2}$, thus resulting in an adiabatic annealing protocol
with the same complexity as Grover's
algorithm~\cite{Grover,Roland.Cerf}. This is in agreement with a
general statement according to which, in terms of complexity,
adiabatic and quantum gate-based protocols are
equivalent~\cite{Aharonov}.

In their basic definitions, both adiabatic annealing and quantum
gate-based approaches require the system to be isolated and at $T=0$.
Of course, such ideal conditions are never satisfied and, although
they are often tacitly assumed and/or supposed to be tackled by more
and more robust and scalable technologies, they still represent a
severe obstacle to the advance of real quantum computers. For
annealers, the ideal $T=0$ adiabatic protocol remains even ill defined
when the time $\tau$ is not sufficiently long as excited states get
populated, which implies heating, contrary to the initial assumption.

There have been attempts to extend adiabatic annealing to open systems
at finite temperature.  On the one hand, we have the result that open
systems follow the instantaneous steady state of the associated
Liouvillian provided that an adiabatic condition similar to that for
isolated systems but with the gap of the Hamiltonian substituted by
the gap of the Liouvillian~\cite{Avron2012,Zanardi2016,Joye2022} is
satisfied.  On the other hand, it is clear that out of equilibrium
annealing offers a plethora of new possibilities, e.g., paths
different from the adiabatic ones, which may provide advantages with
respect to an adiabatic
protocol~\cite{Lidar2005,Amin,Venuti2017,Lidar2018,Passarelli2018,
  Passarelli2019, Lidar2021}.  Nevertheless, without a clear
understanding of the thermodynamics of the quantum system, the whole
picture remains incomplete.

Equations~(\ref{TH_QPTR})-(\ref{pcond3}) reverse the fateful role of
first-order QPTs showing the following groundbreaking properties: (i)
if a system admits a finite-temperature condensation QPT, it can work
as a quantum annealer even if at finite temperature, i.e., not
isolated.  (ii) An annealing protocol consists of any path bringing
the system from any point of the normal phase to any point of the
condensed phase. (iii) During the annealing protocol, the system does
not need to stay at canonical equilibrium. (iv) If the system is at
canonical equilibrium in the condensed phase, a measurement of its
state will provide the target state with a probability (called output
probability in~\cite{Lidar2021}) exponentially close to 1 in size
$N$~\cite{comment_GS_V}.

Note that, in total contrast to statement (iv), in gate-based devices,
the larger is the size, the larger is the probability that the device
undergoes unwanted decoherence. Similarly, in quantum annealers
{operating} away from a condensed phase, at $T>0$ the output
probability decreases exponentially with $N$~\cite{Albash2017}.  It is
in this sense that a quantum annealer based on a condensation QPT
might offer a dramatically important advantage, exploiting at best the
collective and spontaneous mechanism of phase transitions at finite
temperature.

As an example application of statements (i)-(iv) above, consider the
Grover model. It can work as an annealer due to its finite temperature
condensation. We can bring it from the normal to the condensed phase
by following any path crossing the critical line given by
Eq.~(\ref{JcT_Grover}) at some $T_c>0$; in doing so, we are also free
to follow out-of-equilibrium paths, provided we subtract from the
system an amount of energy no smaller than the latent heat
$L=N\{ K_\mathrm{B}T\ln[2\cosh(\beta\Gamma)]-\Gamma\tanh(\beta
\Gamma)\}|_{T=T_c}$, easily derived from $F_\mathrm{cond}$ and
$F_\mathrm{norm}$. Once the model is in the condensed phase, a readout
of the $N$ spins of the model provides the target state with
probability 1, up to corrections exponentially small in $N$.

Regrettably, the Grover Hamiltonian describes an abstract model.  In
fact, its potential $V$, which is totally unstructured, consists of
the sum of two- to $N$-body interactions, a rather unphysical feature.
On the other hand, the fermionic system in an inhomogeneous lattice
could be implemented as a realistic quantum annealer.  In the 1D
lattice that we analyzed, however, the potential $V$ is trivially
structured, allowing for an efficient classical search of its GS (all
fermions in the $N/2$ consecutive attractive sites).  Non trivial
target states requiring an actual quantum search would be obtained by
considering more general lattices, e.g., 1D lattices with attractive
sites arranged in non trivial ways, and/or taking into account
electron interaction. For all these systems a condensation QPT is
still expected to manifest.

\section{Conclusions}%
We have extended to finite temperature a wide class of QPTs
characterized by a condensation in the space of quantum
states. Whereas a rigorous proof of these transitions was provided in
\cite{TH_QPT_proof} (see also Appendix~\ref{proofs.appendix}), here we
proposed an intuitive criterion based on the straightforward
replacement of the ground state energies with the corresponding free
energies, and checked the criterion in two different systems.

A main feature of this class of QPTs is that at any point of the
condensed phase the canonical equilibrium state coincides with the
system GS (one of the degenerate system GSs) with a probability
exponentially close to 1 in the system size.  We explained in detail
how this feature may represent groundbreaking progress for quantum
annealers; see statements (i)-(iv) expounded in the previous section.

Of course, understanding how long it takes for an annealer based on a
condensation QPT to reach canonical equilibrium in the condensed phase
is crucially important but is out of the scope of the present paper.
The thermalization of Grover's model or similar systems with a thermal
bath represented by blackbody radiation could be tackled within the
theories presented in~\cite{Venuti2017,THERM}.  It could be
advantageous to consider annealers based on weakly short-range
interacting systems for which rigorous bounds on the relaxation times
exist in terms of those of the corresponding noninteracting
systems~\cite{Bertini}.  We look forward to reporting the relative
results.

\vspace{5mm}

\begin{acknowledgments}
  We are grateful to A. Polimeni for helpful discussions on the
  experimental implementation of our fermionic model.  M.~O. thanks
  CNPq for funding (Grant No. 307622/2018-5).  M.~O. thanks the
  Istituto Nazionale di Fisica Nucleare, Sezione di Roma 1, and the
  Department of Physics of Sapienza University of Rome for financial
  support and hospitality.
\end{acknowledgments}

\appendix

\section{Proof of Equations.~(\ref{TH_QPTR}) and (\ref{pcond3})}%
\label{proofs.appendix}
For any partition
$\mathcal{H}=\mathcal{H}_\mathrm{cond} \oplus
\mathcal{H}_\mathrm{norm}$, we can prove (see
Ref.~\cite{TH_QPT_proof}) that for any
$\ket{\bm{n}}\in\mathcal{H}_{X}$ ($X$ stands for either
$\mathrm{cond}$ or $\mathrm{norm}$ and $Y$ stands for its complement
$\mathrm{norm}$ or $\mathrm{cond}$)
\begin{align}
  \label{MATRIX}
  &1 \leq \frac{\braket{\bm{n}}{e^{-\beta H}}{\bm{n}}}
    {\braket{\bm{n}}{e^{-\beta H_{X}}}{\bm{n}}}
    \leq
    e^{\beta \Gamma \min\{A_{X}^{(\mathrm{out})},
    A_{Y}^{(\mathrm{out})}\}},
\end{align}
where
$A_{X}^{(\mathrm{out})} = \sup_{\ket{\bm{n}}\in\mathcal{H}_{X}}
\sum_{\ket{\bm{n}'}\in\mathcal{H}_\mathrm{Y}}
|\braket{\bm{n}}{K}{\bm{n}'}|$ represents the maximum number of
outgoing links (nonzero matrix elements of $K=H-V$) from
$\mathcal{H}_{X}$ to $\mathcal{H}_{Y}$.  The proof of
Eq.~(\ref{MATRIX}) is based on the exact probabilistic representation
of the quantum evolution introduced in~\cite{BPDAJL} used at an
imaginary time which is identified with the inverse temperature
$\beta$.  The term
$\min\{A_{X}^{(\mathrm{out})},A_{Y}^{(\mathrm{out})}\}\Gamma$
represents the rate of convergence to 1 of the probability for
crossing the boundary between $\mathcal{H}_{X}$ and $\mathcal{H}_{Y}$
as realized during an infinitely long evolution dictated by the
Hamiltonian $H$. From Eq.~(\ref{MATRIX}) we have (note that a harmless
typo occurred in Eq. (18) of Ref.~\cite{TH_QPT_proof}, where the term
$-{\ln(2)}/{\beta}$ was missed on the right-hand side of that
equation)
\begin{align}
  \label{upper}
  F\leq \min\{F_\mathrm{cond},F_\mathrm{norm}\},
\end{align}
\begin{align}
  \label{lower}
  &F\geq \min\{F_\mathrm{cond},F_\mathrm{norm}\}
    \nonumber \\ &\qquad-\min\{A_{\mathrm{cond}}^{(\mathrm{out})},
                   A_{\mathrm{norm}}^{(\mathrm{out})}\}\Gamma
                   -\frac{\ln(2)}{\beta}.
\end{align}
Equations~(\ref{upper}) and (\ref{lower}) can be derived from
(\ref{MATRIX}) as follows.

By using the left inequality of Eq.~(\ref{MATRIX}) we have
$\braket{\bm{n}}{e^{-\beta H}}{\bm{n}} \geq \braket{\bm{n}}{e^{-\beta
    H_{X}}}{\bm{n}}$, $\forall \ket{\bm{n}}\in\mathcal{H}_{X}$, and
therefore
\begin{align*}
  &\sum_{\ket{\bm{n}}\in\mathcal{H}} \braket{\bm{n}}{e^{-\beta H}}{\bm{n}}
  \\
  &\qquad
    =
    \sum_{\ket{\bm{n}}\in\mathcal{H}_{X}} \braket{\bm{n}}{e^{-\beta H}}{\bm{n}}
    +
    \sum_{\ket{\bm{n}}\in\mathcal{H}_{Y}} \braket{\bm{n}}{e^{-\beta H}}{\bm{n}}
  \\
  &\qquad
    \geq
    \sum_{\ket{\bm{n}}\in\mathcal{H}_{X}}
    \braket{\bm{n}}{e^{-\beta H_X}}{\bm{n}}
    +
    \sum_{\ket{\bm{n}}\in\mathcal{H}_{Y}}
    \braket{\bm{n}}{e^{-\beta H_Y}}{\bm{n}}
\end{align*}
which means
\begin{align*}
  e^{-\beta F} \geq e^{-\beta F_\mathrm{cond}} +
  e^{-\beta F_\mathrm{norm}}\geq e^{-\beta
  \min( F_\mathrm{cond}, F_\mathrm{norm} )},
\end{align*}
equivalent to Eq.~(\ref{upper}).

To prove Eq.~(\ref{lower}) we start from the right inequality of
Eq.~(\ref{MATRIX}), namely,
$\braket{\bm{n}}{e^{-\beta H}}{\bm{n}} \leq \braket{\bm{n}}{e^{-\beta
    H_{X}}}{\bm{n}} e^{\beta \Gamma \min\{A_{X}^{(\mathrm{out})},
  A_{Y}^{(\mathrm{out})}\}}$,
$\forall \ket{\bm{n}}\in\mathcal{H}_{X}$, and obtain
\begin{align*}
  &\sum_{\ket{\bm{n}}\in\mathcal{H}} \braket{\bm{n}}{e^{-\beta H}}{\bm{n}}
  \\
  &\qquad\leq
    \left( \sum_{\ket{\bm{n}}\in\mathcal{H}_{X}}
    \braket{\bm{n}}{e^{-\beta H_X}}{\bm{n}}
    +
    \sum_{\ket{\bm{n}}\in\mathcal{H}_{Y}}
    \braket{\bm{n}}{e^{-\beta H_Y}}{\bm{n}} \right)
  \\
  &\qquad\qquad\times
    e^{\beta \Gamma \min\{A_{X}^{(\mathrm{out})},
    A_{Y}^{(\mathrm{out})}\}},
\end{align*}
which means
\begin{align*}
  e^{-\beta F}
  &\leq
    \left( e^{-\beta F_\mathrm{cond}} + e^{-\beta F_\mathrm{norm}} \right)
    e^{\beta \Gamma \min\{A_{\mathrm{cond}}^{(\mathrm{out})},
    A_{\mathrm{norm}}^{(\mathrm{out})}\}} 
  \\
  &\leq
    2e^{-\beta \min( F_\mathrm{cond}, F_\mathrm{norm} )}
    e^{\beta \Gamma \min\{A_{\mathrm{cond}}^{(\mathrm{out})},
    A_{\mathrm{norm}}^{(\mathrm{out})}\}} ,
\end{align*}
equivalent to Eq.~(\ref{lower}).

Combining Eqs.~(\ref{upper}) and (\ref{lower}) with the assumptions
that $F_\mathrm{cond}$ and $F_\mathrm{norm}$ are extensive in $N$ and
that
$\min\{A_{\mathrm{cond}}^{(\mathrm{out})},
A_{\mathrm{norm}}^{(\mathrm{out})}\}=\mathrm{o}(N)$ proves that
Eqs.~(\ref{TH_QPTR}) hold true up to terms becoming negligible in the
TDL.  In our setting, according to the definition of the subspace
$\mathcal{H}_\mathrm{cond}$, we always have
$\min\{A_{\mathrm{cond}}^{(\mathrm{out})},
A_{\mathrm{norm}}^{(\mathrm{out})}\}=
A_{\mathrm{norm}}^{(\mathrm{out})}=\mathrm{o}(N)$.  This represents a
reasonably general property. In the Grover model, for instance, we
have $A_{\mathrm{norm}}^{(\mathrm{out})}=1$, while
$A_{\mathrm{cond}}^{(\mathrm{out})}=N$. Similarly, it is easy to check
that in the model of free fermions in an inhomogeneous lattice, with
the choice $N_p=N_i=N/2$ we have
$A_{\mathrm{norm}}^{(\mathrm{out})}=2$, while
$A_{\mathrm{cond}}^{(\mathrm{out})}=N/2$.  The important point is
that, in most of the systems of interest, the conditions
$M_{\mathrm{cond}}/M\to 0$ and
$A_{\mathrm{norm}}^{(\mathrm{out})}/N\to 0$ are
equivalent~\cite{TH_QPT_proof} and, under any of these conditions,
Eqs.~(\ref{upper}) and (\ref{lower}), up to $\mathrm{o}(N)$ terms,
provide Eqs.~(\ref{TH_QPTR}), namely, the generalization to finite
temperature of the condensation QPTs, as suggested by the natural
criterion of substituting GS energies with free energies.

To prove Eq.~(\ref{pcond3}), we start inserting the expression of the
Gibbs state $\rho=e^{-\beta H}/\tr e^{-\beta H}$ into the definition
of $p_\mathrm{cond}$,
\begin{align}
  \label{SM_pcond3}
  p_\mathrm{cond}
  &=
    \sum_{\ket{\bm{n}}\in\mathcal{H}_\mathrm{cond}}
    \braket{\bm{n}}{\rho}{\bm{n}}
    \nonumber\\&=
  \frac{\sum_{\ket{\bm{n}}\in\mathcal{H}_\mathrm{cond}}
  \braket{\bm{n}}{e^{-\beta H}}{\bm{n}}}
  {\sum_{\ket{\bm{n}}\in\mathcal{H}}
  \braket{\bm{n}}{e^{-\beta H}}{\bm{n}}}
  \nonumber \\
  &=
    \frac{1}{1+
    \frac{\sum_{\ket{\bm{n}}\in\mathcal{H}_\mathrm{norm}}
    \braket{\bm{n}}{e^{-\beta H}}{\bm{n}}}
    {\sum_{\ket{\bm{n}}\in\mathcal{H}_\mathrm{cond}}
    \braket{\bm{n}}{e^{-\beta H}}{\bm{n}}}}.
\end{align}
By using again the left and right inequalities of Eq.~(\ref{MATRIX}),
we have the following inequalities, assuming
$\min\{A_{\mathrm{cond}}^{(\mathrm{out})},
A_{\mathrm{norm}}^{(\mathrm{out})}\}=A_{\mathrm{norm}}^{(\mathrm{out})}$:
\begin{align}
  \label{SM_pcond3a}
  p_\mathrm{cond}
  &\leq
    \frac{1}{1+
    \frac{\sum_{\ket{\bm{n}}\in\mathcal{H}_\mathrm{norm}}
    \braket{\bm{n}}{e^{-\beta H_\mathrm{norm}}}{\bm{n}}}
    {\sum_{\ket{\bm{n}}\in\mathcal{H}_\mathrm{cond}}
    \braket{\bm{n}}{e^{-\beta H}}{\bm{n}}}}
    \nonumber \\
  &\leq
    \frac{1}{1+
    \frac{\sum_{\ket{\bm{n}}\in\mathcal{H}_\mathrm{norm}}
    \braket{\bm{n}}{e^{-\beta H_\mathrm{norm}}}{\bm{n}}}
    {\sum_{\ket{\bm{n}}\in\mathcal{H}_\mathrm{cond}}
    e^{\beta \Gamma A_{\mathrm{norm}}^{(\mathrm{out})}}
    \braket{\bm{n}}{e^{-\beta H_\mathrm{cond}}}{\bm{n}}}},
\end{align}
and
\begin{align}
  \label{SM_pcond3b}
  p_\mathrm{cond}
  &\geq
    \frac{1}{1+
    \frac{\sum_{\ket{\bm{n}}\in\mathcal{H}_\mathrm{norm}}
    e^{\beta \Gamma A_{\mathrm{norm}}^{(\mathrm{out})}}\braket{\bm{n}}
    {e^{-\beta H_\mathrm{norm}}}{\bm{n}}}
    {\sum_{\ket{\bm{n}}\in\mathcal{H}_\mathrm{cond}}
    \braket{\bm{n}}{e^{-\beta H}}{\bm{n}}}}
    \nonumber \\
  &\geq
    \frac{1}{1+
    \frac{\sum_{\ket{\bm{n}}\in\mathcal{H}_\mathrm{norm}}
    e^{\beta \Gamma A_{\mathrm{norm}}^{(\mathrm{out})}}
    \braket{\bm{n}}{e^{-\beta H_\mathrm{norm}}}{\bm{n}}}
    {\sum_{\ket{\bm{n}}\in\mathcal{H}_\mathrm{cond}}
    \braket{\bm{n}}{e^{-\beta H_\mathrm{cond}}}{\bm{n}}}}.
\end{align}
On making use of the definitions of $F_{\mathrm{cond}}$ and
$F_{\mathrm{norm}}$, the two inequalities above can be rewritten as
\begin{align*}
  \label{SM_pcond4}
  &p_\mathrm{cond}\leq
    \frac{1}{1+e^{-\beta (F_{\mathrm{norm}}-F_{\mathrm{cond}}) -
    \beta \Gamma A_{\mathrm{norm}}^{(\mathrm{out})}} },
  \\
  &p_\mathrm{cond} \geq
    \frac{1}{1+e^{-\beta (F_{\mathrm{norm}}-F_{\mathrm{cond}})
    + \beta \Gamma A_{\mathrm{norm}}^{(\mathrm{out})}} },
\end{align*}
which, on assuming again that
$A_{\mathrm{norm}}^{(\mathrm{out})}=\mathrm{o}(N)$, implies
Eq.~(\ref{pcond3}).

\section{Order parameter of Grover model}%
\begin{figure}[h]
  \begin{center}
    {\includegraphics[height=5cm,clip]{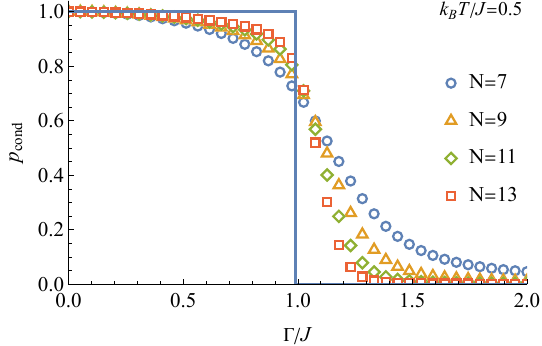}}
    {\includegraphics[height=5cm,clip]{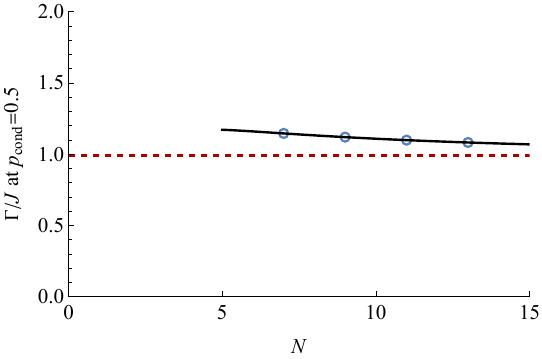}}
    \caption{Top: Order parameter $p_\mathrm{cond}$ versus $\Gamma/J$
      for the Grover model at thermal equilibrium at temperature
      $k_BT/J=0.5$ for $N=7,9,11,13$ (open symbols).  The step-like
      solid line is the asymptotic $N\to\infty$ exact value.  Bottom:
      Value of $\Gamma/J$ at $p_\mathrm{cond}=1/2$ for different $N$
      (open circles).  The function $a+b/N+c/N^2$ fits quite well the
      data (black solid line, $a=0.96$, $b=1.88$, and $c=-4.18$) and
      predicts the asymptotic exact value $\Gammac/J=0.99$ (dashed red
      line) with a $3\%$ error.  }
    \label{pcond_Grover_05}
  \end{center}
\end{figure}

\begin{figure}[h]
  \begin{center}
    {\includegraphics[height=5cm,clip]{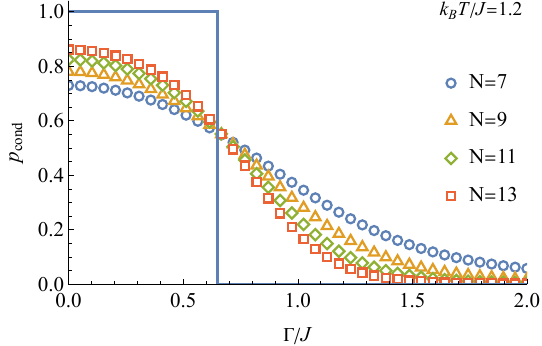}}
    {\includegraphics[height=5cm,clip]{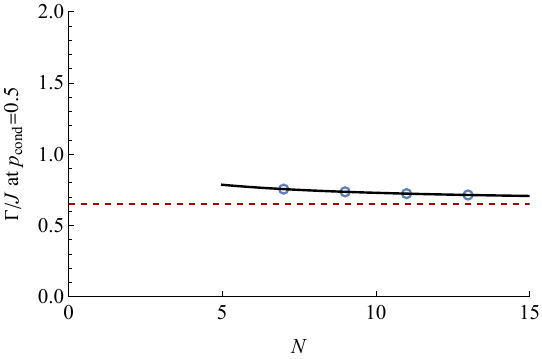}}
    \caption{As in Fig.~\ref{pcond_Grover_05} but at temperature
      $k_BT/J=1.2$.  The fit parameters ($a=0.67$, $b=0.79$, and
      $c=-0.75$) predict the asymptotic exact value $\Gammac/J=0.65$
      with a $3\%$ error.  }
    \label{pcond_Grover_12}
  \end{center}
\end{figure}

\label{Grover.appendix}
Figures \ref{pcond_Grover_05} and \ref{pcond_Grover_12} show the order
parameter $p_\mathrm{cond}$ evaluated numerically in the Grover model
as a function of $\Gamma/J$ for different values of $N$. Note that the
higher the temperature is the slower the convergence of the data to
the $N\to\infty$ limit of $p_\mathrm{cond}$ is, represented by the
step like solid line.  Nevertheless, the fit of $a+b/N+c/N^2$ to the
values of $\Gamma/J$ at $p_\mathrm{cond}=0.5$ provides an asymptotic
value of $\Gammac/J=a$ in good agreement with the exact value.
\vspace{2cm}

\section{Free fermions in a 1D inhomogeneous lattice}%
  
\subsection{$T=0$}%
\label{fermions.appendix.1}
First of all, we report the exact numerical analysis of the energy
density $E/\Np$ and its derivative $d(E/\Np)/dg$, as well as of the
order parameter $p_\mathrm{cond}$ at $T=0$.  The zero-temperature case
is particularly simple; we do not need to evaluate awkward canonical
partition functions as in the case with $T>0$ and we can consider
systems of large size, with behavior practically indistinguishable
from their TDLs.
  
\begin{figure}
  \begin{center}
    {\includegraphics[width=\columnwidth,clip]
      {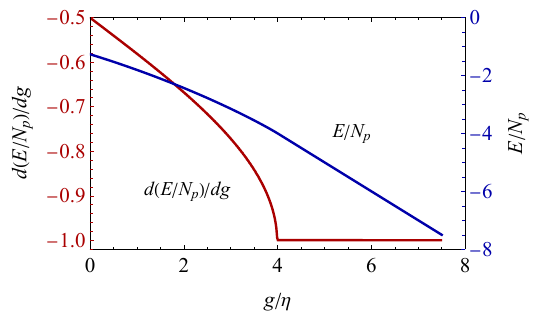}}
    \caption{Ground-state energy per particle $E/\Np$ (blue line,
      right scale) and its first derivative with respect to the
      parameter $g$ (red line, left scale) as a function of $g/\eta$
      for the system of free fermions in a 1D inhomogeneous lattice
      described by the Hamiltonian~(\ref{FermionH}) with
      $\Np=\Nimp=N/2=133$. For this value of $\Np$ both curves have
      practically reached their TDL. }
    \label{new2}
  \end{center}
\end{figure}

Figure \ref{new2} clearly shows the presence of a point of non
analyticity at $\gc=4$ in the energy density of the system. This
proves the existence of a QPT.

\begin{figure}
  \begin{center}
    {\includegraphics[width=\columnwidth,clip]
      {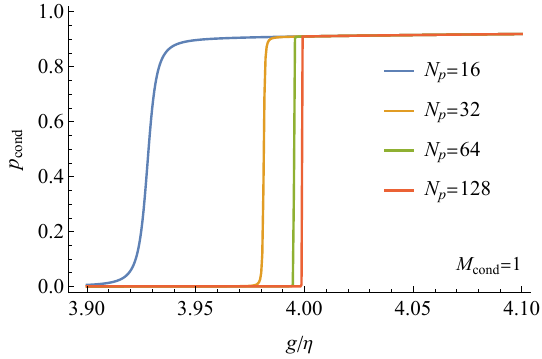}}
    {\includegraphics[width=\columnwidth,clip]
      {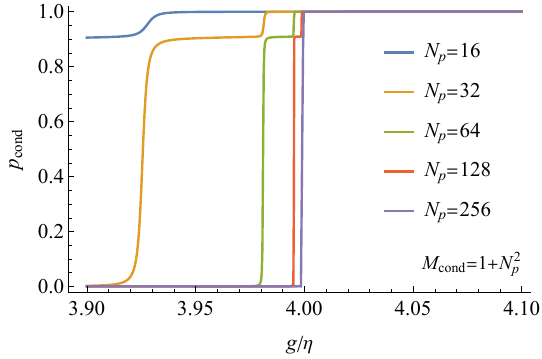}}
    \caption{Order parameter $p_\mathrm{cond}$ versus $g/\eta$ at
      $T=0$ for the system of free fermions in a 1D inhomogeneous
      lattice described by the Hamiltonian~(\ref{FermionH}) with
      $\Np=\Nimp=N/2=16,32,64,128,256$. The top panel corresponds to
      the choice $M_\mathrm{cond} = 1$, while in the bottom panel we
      have $M_\mathrm{cond} = 1+\Np^2$.  In both panels
      $p_\mathrm{cond}$ has practically reached its TDL for the
      largest $\Np$ shown.}
    \label{new1}
  \end{center}
\end{figure}

Of course, the above result holds regardless of the choice of
$M_\mathrm{cond}$. In fact, in our approach any suitable choice of
$M_\mathrm{cond}$, e.g., $M_\mathrm{cond}=1$ or
$M_\mathrm{cond}=1+\Np^2$, will lead to the same critical point $\gc$.
However, as shown in Fig.~\ref{new1}, only in the latter case do we
have a complete jump of the order parameter $p_\mathrm{cond}$ between
0 and 1 at $\gc=4$.  According to Eq.~(\ref{pcond3}), a 0-1 jump of
$p_\mathrm{cond}$ at $g=\gc$ corresponds, for $T\to 0$, to a true
crossing of the GS energies $E_\mathrm{cond}$ and $E_\mathrm{norm}$ at
$g=\gc$ (more correctly, a true crossing of the TDL of the GS energy
densities $E_\mathrm{cond}/\Np$ and $E_\mathrm{norm}/\Np$). On the
other hand, for $M_\mathrm{cond}=1$ the two energies $E_\mathrm{cond}$
and $E_\mathrm{norm}$, which are distinct for $g<\gc$ (because
$p_\mathrm{cond}=0$ and therefore
$E_\mathrm{norm} < E_\mathrm{cond}$), merge and remain identical for
$g\geq \gc$ (because $0<p_\mathrm{cond}<1$). Note that for
$M_\mathrm{cond}=1$ the order parameter slowly increases for $g>\gc$
and approaches 1 for $g\to\infty$. In this limit the GS $\ket{E}$
coincides with the lowest eigenstate of $V$, i.e.,
$\ket{E} \in \mathcal{H}_\mathrm{cond}$. On the other hand, for any
finite value $g\ge\gc$, $\ket{E}$ has components in both
$\mathcal{H}_\mathrm{cond}$ and $\mathcal{H}_\mathrm{norm}$, with a
net prevalence in the former. More precisely, the probability for the
system to be found in $\mathcal{H}_\mathrm{cond}$ (i.e., in the lowest
eigenstate of $V$) turns out to be larger than about 90\%
for~$g\gtrsim \gc$.

It is interesting to observe the following technical aspect of the
present QPT.  From Fig.~\ref{new2} we see that it is the second
derivative of the energy density that diverges at $g=\gc$, whereas
from Fig.~\ref{new1} we see that the order parameter undergoes a
finite jump at $g=\gc$. In other words, in this peculiar model, the
QPT turns out to be a hybrid one which has a second-order nature (more
precisely, it is a second-order transition within the so-called
``$\lambda$-transition'' class; see, for example
Ref.~\cite{Fidelity}), when seen with respect to the GS energy but a
first-order nature when seen with respect to the order parameter.

\subsection{$T>0$}%
\begin{figure}
  \begin{center}
    {\includegraphics[width=\columnwidth,clip]
      {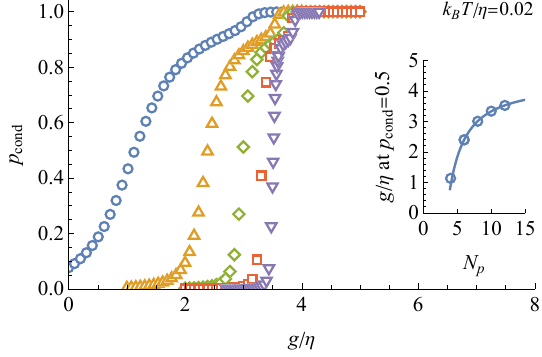}}
    \caption{Order parameter $p_\mathrm{cond}$ versus $g/\eta$ for the
      fermionic system~(\ref{FermionH}) with
      $\Np=\Nimp=N/2=4,6,8,10,12$ (symbols from left to right).  We
      put $M_\mathrm{cond} = 1+\Np^2$ and the system is at canonical
      equilibrium at temperature $k_BT/\eta=0.02$. Inset: value of
      $g/\eta$ at $p_\mathrm{cond}=1/2$ for different $\Np$.  Data
      (circles) compare quite well with the fitting function
      $a+b/\Np+c/\Np^2$ (solid line, $a=4.277$, $b=-6.860$ and
      $c=-26.841$).  }
    \label{pcond_fermions_impurities_002}
  \end{center}
\end{figure}

\begin{figure}
  \begin{center}
    {\includegraphics[width=\columnwidth,clip]
      {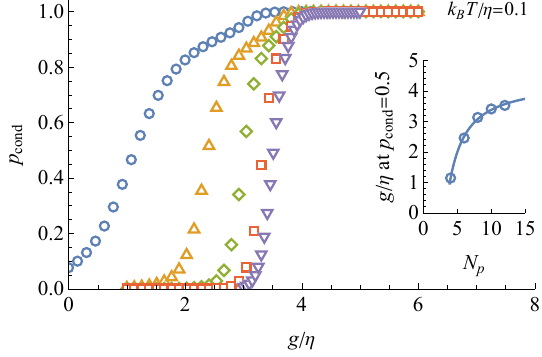}}
    \caption{ As in Fig. \ref{pcond_fermions_impurities_002}, but at
      temperature $k_BT/\eta=0.1$.  The fit gives $a=4.303$,
      $b=-7.058$ and $c=-22.568$.  }
    \label{pcond_fermions_impurities_01}
  \end{center}
\end{figure}

\begin{figure}
  \begin{center}
    {\includegraphics[width=\columnwidth,clip]
      {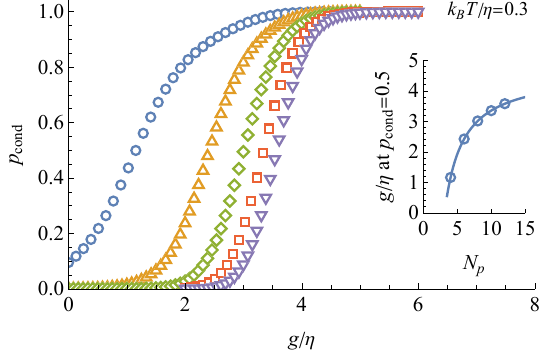}}
    \caption{ As in Fig. \ref{pcond_fermions_impurities_002}, but at
      temperature $k_BT/\eta=0.3$.  The fit gives $a=4.622$,
      $b=-11.987$ and $c=-7.506$.  }
    \label{pcond_fermions_impurities_03}
  \end{center}
\end{figure}

\begin{figure}
  \begin{center}
    {\includegraphics[width=\columnwidth,clip]
      {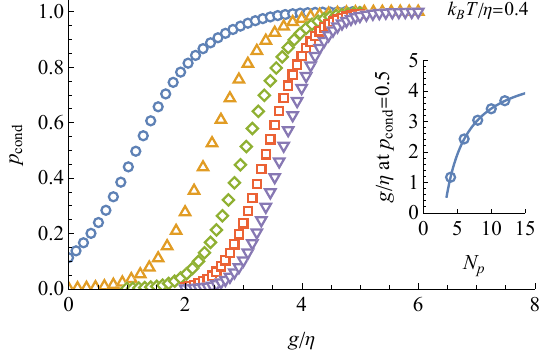}}
    \caption{ As in Fig. \ref{pcond_fermions_impurities_002}, but at
      temperature $k_BT/\eta=0.4$.  The fit gives $a=4.922$,
      $b=-15.144$ and $c=0.4423$.  }
    \label{pcond_fermions_impurities_04}
  \end{center}
\end{figure}

\begin{figure}
  \begin{center}
    {\includegraphics[width=\columnwidth,clip]
      {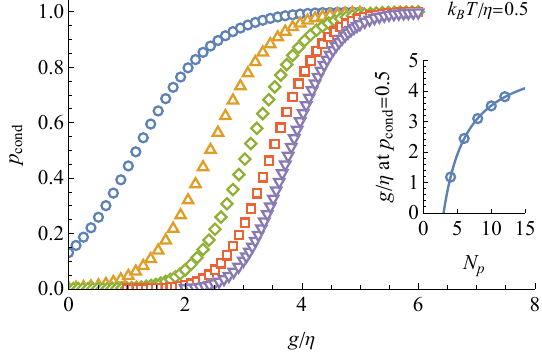}}
    \caption{ As in Fig. \ref{pcond_fermions_impurities_002}, but at
      temperature $k_BT/\eta=0.5$.  The fit gives $a=5.277$,
      $b=-18.517$ and $c=8.364$.  }
    \label{pcond_fermions_impurities_05}
  \end{center}
\end{figure}

\begin{figure}
  \begin{center}
    {\includegraphics[width=\columnwidth,clip]
      {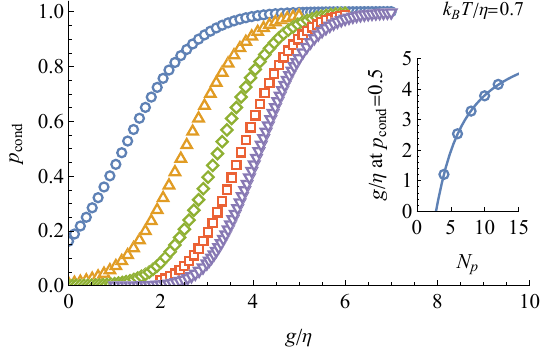}}
    \caption{ As in Fig. \ref{pcond_fermions_impurities_002}, but at
      temperature $k_BT/\eta=0.7$.  The fit gives $a=6.060$,
      $b=-24.931$ and $c=21.991$.  }
    \label{pcond_fermions_impurities_07}
  \end{center}
\end{figure}

\begin{figure}
  \begin{center}
    {\includegraphics[width=\columnwidth,clip]
      {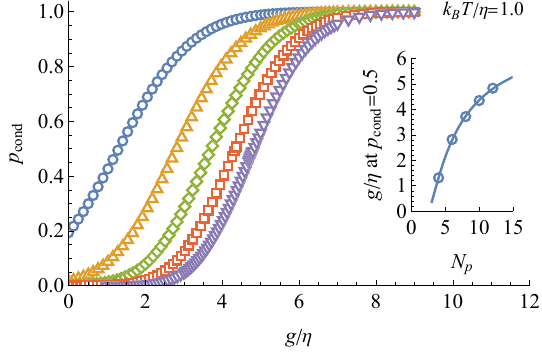}}
    \caption{ As in Fig. \ref{pcond_fermions_impurities_002}, but at
      temperature $k_BT/\eta=1.0$.  The fit gives $a=7.359$,
      $b=-33815$ and $c=38.458$.  }
    \label{pcond_fermions_impurities_10}
  \end{center}
\end{figure}

\begin{figure}
  \begin{center}
    {\includegraphics[width=\columnwidth,clip]
      {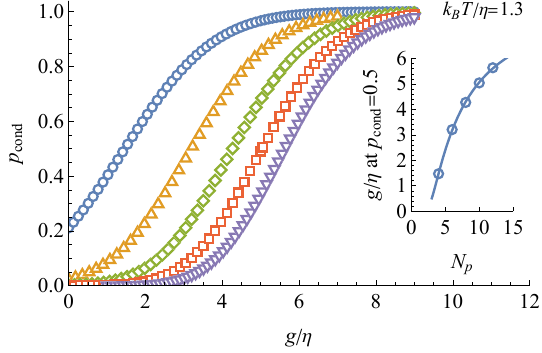}}
    \caption{ As in Fig. \ref{pcond_fermions_impurities_002}, but at
      temperature $k_BT/\eta=1.3$.  The fit gives $a=8.774$,
      $b=-42.407$ and $c=52.764$.  }
    \label{pcond_fermions_impurities_13}
  \end{center}
\end{figure}

\label{fermions.appendix.2}
In
Figs.~\ref{pcond_fermions_impurities_002}-\ref{pcond_fermions_impurities_13},
we report the values of $p_\mathrm{cond}$ obtained numerically for the
system (\ref{FermionH}) at canonical equilibrium at different
temperatures $T$. In each plot, the parameter $a$ of the fit is taken
as the critical value $\gc(T)$ shown in
Fig.~\ref{phase_diagram_fermions_impurities}.

\vspace{2cm}

\begin{thebibliography}{15}
  

\bibitem{SGCS} S.~L.~Sondhi, S.~M.~Girvin, J.~P.~Carini, and
  D.~Shahar, Continuous quantum phase transitions,
  Rev. Mod. Phys. \textbf{69}, 315 (1997).
  
\bibitem{KB} T.~R.~Kirkpatrick and D.~Belitz, Quantum phase
  transitions in electronic systems, in \textit{Electron Correlations
    in the Solid State}, edited by N.~H.~March, (Imperial College
  Press, London 1999), pp. 297-–370.

\bibitem{Vojta} T.~Vojta, Quantum phase transitions in electronic
  systems, Ann. Phys. (N.Y.) \textbf{9}, 403 (2000).

\bibitem{Sachdev} S.~Sachdev, \textit{Quantum Phase Transitions}
  (Cambridge University Press, Cambridge 2000).

\bibitem{Zanardi2007} P.~Zanardi, H.~T.~Quan, X.~Wang, and C.~P.~Sun,
  Mixed-state fidelity and quantum criticality at finite temperature,
  Phys. Rev. A \textbf{75}, 032109 (2007).

\bibitem{Fidelity} H.~T.~Quan and F.~M.~Cucchietti, Quantum fidelity
  and thermal phase transitions, Phys. Rev. E \textbf{79}, 031101
  (2009).
  
\bibitem{Carr} L.~D.~Carr, Understanding Quantum Phase Transitions,
  (CRC Press, Boca Raton, FL, 2010).

\bibitem{Plastino} A.~Plastino and E.~M.~F.~Curado, Finite temperature
  approach to quantum phase transitions, International Journal of
  Bifurcation and Chaos \textbf{20}, 397 (2010).
  
\bibitem{Sebenik} A.~B.~Finilla, M.~A.~Gomez, C.~Sebenik, and
  D.~J.~Doll, Quantum annealing: A new method for minimizing
  multidimensional functions, Chem. Phys. Lett. \textbf{219}, 343
  (1994).

\bibitem{Nishimori} T.~Kadowaki and H.~Nishimori, Quantum annealing in
  the transverse Ising model, Phys. Rev. E \textbf{58}, 5355 (1998).
    
\bibitem{Santoro} G.~E.~Santoro and E.~Tosatti, Optimization using
  quantum mechanics: quantum annealing through adiabatic evolution,
  J. Phys. A \textbf{39}, R393 (2006).

\bibitem{QPTA} M.~Ostilli and C.~Presilla, First-order quantum phase
  transitions as condensations in the space of states, J. Phys. A:
  Math. Theor. \textbf{54}, 055005 (2021).

\bibitem{BPDAJL} M.~Beccaria, C.~Presilla, G.~F.~De Angelis, and
  G.~Jona-Lasinio, An exact representation of the fermion dynamics in
  terms of Poisson processes and its connection with Monte Carlo
  algorithms, Europhys. Lett. \textbf{48}, 243 (1999).
 
\bibitem{TH_QPT_proof} M.~Ostilli and C.~Presilla, Finite temperature
  quantum condensations in the space of states: general proof,
  J. Phys. A: Math. Theor. \textbf{55}, 505004 (2022).

\bibitem{Continentino} M.~A.~Continentino and A.~S.~Ferreira,
  First-order quantum phase transitions, J. Magnetism and Magnetic
  Materials \textbf{310}, 828 (2007).

\bibitem{Pelissetto} M.~Campostrini, J.~Nespolo, A.~Pelissetto, and
  E.~Vicari, Finite-Size Scaling at First-Order Quantum Transitions
  Phys. Rev. Lett. \textbf{113}, 070402 (2014).
  
\bibitem{WC_QPT} M.~Ostilli and C.~Presilla, Wigner crystallization of
  electrons in a one-dimensional lattice: a condensation in the space
  of states, Phys. Rev. Lett. \textbf{127}, 040601 (2021).

\bibitem{Grover} L.~K.~Grover, A fast quantum-mechanical search
  algorithm for database search, in \textit{Proceedings, 28th Annual
    ACM Symposium on the Theory of Computing (STOC)} (ACM Press, New
  York, Philadelphia, Pennsylvania, 1996), pp.  212--219; Quantum
  Mechanics Helps in Searching for a Needle in a Haystack,
  Phys. Rev. Lett. \textbf{79}, 325 (1997); From Schr\"odinger's
  equation to the quantum search algorithm, Am. J. Phys. \textbf{69},
  769 (2001).
  
\bibitem{Roland.Cerf} J.~Roland and N.~J.~Cerf, Quantum search by
  local adiabatic evolution, Phys. Rev. A, \textbf{65}, 042308 (2002).

\bibitem{Farhi.Goldstone} E. Farhi, J. Goldstone, S. Gutmann, and
  D. Nagaj, How to make the quantum adiabatic algorithm fail,
  Int. J. Quantum. Inf. \textbf{6}, 503–516 (2008).
  
\bibitem{Jorg:2008} T.~J\"org, F.~Krzakala, J.~Kurchan, and
  A.~C.~Maggs, Simple Glass Models and Their Quantum Annealing,
  Phys. Rev. Lett. \textbf{101}, 147204 (2008).

\bibitem{Jorg:2010} T.~J\"org, F.~Krzakala, J.~Kurchan, A.~C.~Maggs,
  and J.~Pujos, Energy gaps in quantum first-order mean-field–like
  transitions: The problems that quantum annealing cannot solve,
  Europhysics Letters \textbf{89}, 40004 (2010).

\bibitem{Blaizot_Ripka} J.~P.~Blaizot and G.~Ripka, \textit{Quantum
    Theory of Finite Systems} (MIT Press, Cambridge, MA, 1986).

\bibitem{2Dheterostructures} Qian~Wang, Lin~Zhang1, Xuejuan~Liu and
  Sha~Li, Two-Dimensional Semiconductor Heterojunctions for
  Optoelectronics and Electronics, Front. Energy Res., 9:802055
  (2021).

\bibitem{nanowires} O.~Arif, V.~Zannier, F.~Rossi, D.~De Matteis, K.~
  Kress, M. De Luca, I.~Zardo and L.~Sorba, GaAs/GaP superlattice
  nanowires: growth, vibrational and optical properties, Nanoscale
  \textbf{15}, 1145 (2023).

\bibitem{Polimeni} D.~Tedeschi, M.~De Luca, A.~Polimeni,
  Photoluminescence spectroscopy applied to semiconducting nanowires:
  A valuable probe for assessing lattice defects, crystal structures,
  and carriers' temperature, in \textit{Fundamental Properties of
    Semiconductor Nanowires}, edited by N.~Fukata and R.~Riccardo
  (Springer, Singapore, 2021), pp. 289--306.
  
\bibitem{Dwave} T.~Lanting \textit{et al.}, Entanglement in a Quantum
  Annealing Processor, Phys. Rev. X \textbf{4}, 021041 (2014).

\bibitem{Kato} T.~Kato, On the adiabatic theorem of quantum mechanics,
  J. Phys. Soc. Jpn. \textbf{5}, 435 (1950).

\bibitem{Farhi.Gutmann} E.~Farhi and S.~Gutmann, Analog analogue of a
  digital quantum computation, Phys. Rev. A \textbf{57}, 2403 (1998).
  
\bibitem{Warzel2015} J.~Adame and S.~Warzel, Exponential vanishing of
  the ground-state gap of the quantum random energy model via
  adiabatic quantum computing, Journal of Mathematical Physics
  \textbf{56}, 113301 (2015).

\bibitem{Altshuler} B.~Altshuler, H.~Krovi, and J.~Roland, Anderson
  localization makes adiabatic quantum optimization fail, Proc. Natl
  Acad. Sci. USA \textbf{107}, 12446 (2010).
 
\bibitem{ANN} C.~Presilla and M.~Ostilli, Phase transitions and gaps
  in quantum random energy models, Phys. A (Amsterdam, Neth.)
  \textbf{515}, 57 (2019).
    
\bibitem{Aharonov} D.~Aharonov, W.~van~Dam, J.~Kempe, Z.~Landau,
  S.~Lloyd, and O.~Regev, Adiabatic quantum computation is equivalent
  to standard quantum computation SIAM Review \textbf{50}, 755–787
  (2008).

\bibitem{Avron2012} J.~E.~Avron, M.~Fraas, G.~M.~Graf, and P.~Grech,
  Adiabatic Theorems for Generators of Contracting Evolutions,
  Commun. Math. Phys. \textbf{314}, 163 (2012).
    
\bibitem{Zanardi2016} L.~Campos Venuti, T.~Albash, D.~A.~Lidar, and
  P.~Zanardi, Adiabaticity in open quantum systems, Phys. Rev. A
  \textbf{93}, 032118 (2016).

\bibitem{Joye2022} A.~Joye, Adiabatic Lindbladian Evolution with Small
  Dissipators, Commun. Math. Phys. \textbf{391}, 223 (2022).

\bibitem{Lidar2005} M. S. Sarandy and D. A. Lidar, Adiabatic
  approximation in open quantum systems, Phys. Rev. A \textbf{71},
  012331 (2005).
  
\bibitem{Amin} M. H. S. Amin, Peter J. Love, and C. J. S. Truncik,
  Thermally Assisted Adiabatic Quantum Computation,
  Phys. Rev. Lett. \textbf{100}, 060503 (2008).

\bibitem{Venuti2017} L. Campos Venuti, T. Albash, M. Marvian,
  D. Lidar, and P. Zanardi, Relaxation versus adiabatic quantum
  steady-state preparation, Phys. Rev. A \textbf{95}, 042302 (2017).
    
\bibitem{Lidar2018} A.~Mishra, T.~Albash, and D.~A.~Lidar, Finite
  temperature quantum annealing solving exponentially small gap
  problem with non-monotonic success probability,
  Nat. Commun. \textbf{9}, 2917 (2018).
  
\bibitem{Passarelli2018} G.~Passarelli, G.~De Filippis, V.~Cataudella,
  and P.~Lucignano, Dissipative environment may improve the quantum
  annealing performances of the ferromagnetic $p$-spin model,
  Phys. Rev. A \textbf{97}, 022319 (2018).
  
\bibitem{Passarelli2019} G.~Passarelli, V.~Cataudella, and
  P.~Lucignano, Improving quantum annealing of the ferromagnetic
  p-spin model through pausing, Phys. Rev. B \textbf{100}, 024302
  (2019).
  
\bibitem{Lidar2021} E.~J.~Crosson and D.~A.~Lidar, Prospects for
  quantum enhancement with diabatic quantum annealing,
  Nat. Rev. Phys. \textbf{3}, 466 (2021).
  
\bibitem{comment_GS_V} For simplicity, we assumed that the potential
  $V$ has a nondegenerate GS, as in the Grover model or the fermionic
  system.  If the GS of V has degeneracy $d$, statement (iv) needs to
  be rephrased as follows: if the annealer is at canonical equilibrium
  in the condensed phase, a measurement of its state will provide one
  of the $d$ target states with a probability exponentially close to 1
  in the size $N$.
  
\bibitem{Albash2017} T.~Albash, V.~Martin-Mayor, and I.~Hen,
  Temperature Scaling Law for Quantum Annealing Optimizers,
  Phys. Rev. Lett. \textbf{119}, 110502 (2017).
  
\bibitem{THERM} M~Ostilli and C.~Presilla, Thermalization of
  noninteracting quantum systems coupled to blackbody radiation: A
  Lindblad-based analysis, Phys. Rev. A \textbf{95}, 062112 (2017).

\bibitem{Bertini} L.~Bertini, A.~De~Sole, G.~Posta, and C.~Presilla,
  Perturbative criteria for the ergodicity of interacting dissipative
  quantum systems (unpublished).


\end{thebibliography}
\end{document}